\RequirePackage{amsmath}
\documentclass[runningheads]{llncs}
\usepackage[T1]{fontenc}
\usepackage{graphicx}
\usepackage{listings}
\usepackage{xcolor}
\usepackage{flushend} 
\usepackage{tikz}
\usetikzlibrary{shapes.geometric, arrows, positioning, calc}

\definecolor{linkc}{rgb}{0.1,0.1,.8}
\definecolor{darkgreen}{rgb}{0,0.5,0}
\definecolor{midblue}{rgb}{0,0,0.7}
\usepackage[bookmarks=false,colorlinks=true,citecolor=linkc,linkcolor=linkc,filecolor=linkc,urlcolor=linkc]{hyperref}
\usepackage{url}

\usepackage{cleveref}
\usepackage{pgfplots}
\usepackage{makecell}
\usepackage{multirow}
\usepackage{comment}
\usepackage{makecell}
\usepackage{adjustbox}

\pgfplotsset{compat=1.18}

\newcommand{\mcell}[2]{\makecell[lt]{#1\\#2}}
\newcommand{\hhead}[1]{\textbf{#1}}
\newcommand{\vhead}[1]{\rotatebox[origin=l]{90}{\textbf{{\scriptsize #1}}}}
\newcommand{\arxivid}[1]{\href{https:arxiv.org/pdf/#1.pdf}{#1}}
\newcommand{\arxivth}[1]{\href{https:arxiv.org/pdf/hep-th/#1.pdf}{#1}}

\newcommand{\lstFigSize}{\scriptsize}
\newcommand{\lstDisplaySize}{\footnotesize}
\lstdefinestyle{lstFigStyle}{
   xleftmargin=2.5em,
   mathescape=true,
   escapechar=@,
   basicstyle=\lstFigSize\ttfamily,
   keywordstyle=\lstFigSize\color{midblue}\ttfamily\bfseries,
   commentstyle=\lstFigSize\slshape\color{darkgreen},
   numbers=left,
   numberstyle=\lstFigSize\rmfamily
}  
\lstdefinestyle{lstDisplayStyle}{
   xleftmargin=\parindent,
   mathescape=true,
   escapechar=@,
   basicstyle=\lstDisplaySize\ttfamily,
   keywordstyle=\lstDisplaySize\color{midblue}\ttfamily\bfseries,
   commentstyle=\lstDisplaySize\slshape\color{darkgreen},
   numbers=none
}  
\begin{document}
\title{\mbox{Using General Large Language Models to} \mbox{Classify Mathematical Documents}}
\author{Patrick D.F. Ion\inst{1} \and Stephen M. Watt\inst{2}}

\institute{International Mathematical Knowledge Trust [Mathematical Reviews ret'd]
\email{pion@umich.edu}\\[\baselineskip]
\and
Cheriton School of Computer Science, University of Waterloo, Canada\\
\email{smwatt@uwaterloo.ca}}
\maketitle              
\begin{abstract}
In this article we report on an initial exploration to assess the viability of using the general large language models (LLMs), recently made public, to classify mathematical documents. 
Automated classification would be useful from the applied perspective of improving the navigation of the literature and the more open-ended goal of identifying relations among mathematical results.
The Mathematical Subject Classification MSC 2020, from MathSciNet and zbMATH, is widely used and there is a significant corpus of ground truth material in the open literature. 
We have evaluated the classification of preprint articles from \texttt{arXiv.org} according to MSC 2020.
The experiment used only the title and abstract alone --- not the entire paper.  Since this was early in the use of chatbots and the development of their APIs, we report here on what was carried out by hand.  Of course, the automation of the process will have to follow if it is to be generally useful.
We found that in about 60\% of our sample the LLM produced a primary classification matching that already reported on arXiv.
In about half of those instances, there were additional primary classifications that were not detected.
In about 40\% of our sample, the LLM suggested a different classification than what was provided.
A detailed examination of these cases, however, showed that the LLM-suggested classifications were in most cases \textit{better} than those provided.
\end{abstract}
%
%
%
\section{Introduction}
\label{sec:introduction}
There's been much interest in AI created by the advent, in the last year or so, of chatbots based on Large Language Models with their newest technological improvements. It's an obvious thing to see if the classification of mathematics papers can be done using the new publicly offered tools.  Mathematical reviewing services have been categorizing the research in the publications they consider since their inception. 

The online services zbMATH, the successor to Zentralblatt f\"ur
Mathematik und Ihre Anwendungen (Zbl), and MathSciNet (MSN), the product of Mathematical Reviews (MR), jointly publish the Mathematics Subject Classification scheme now in its latest 2020 revision (MSC2020).  Furthermore these services assign MSC classifications to the items they deal with, as do many other scientific sources, including the most influential repository of mathematical preprints, arXiv.  So there's a great deal of ground truth available against which to evaluate the performance of LLM-supported classification.

Our  objective is
to evaluate the effectiveness and reliability of classification performed in an automated manner using the new tools.
At the time of writing, there are already many public tools:  ChatGPT from OpenAI in versions 3.5 and 4, CoPilot from Microsoft.
Gemini (previously Bard) from Google, LLama 2 from Meta (formerly FaceBook), Grok from {X}\.ai (formerly Twitter); in addition to these offerings from tech behemoths, the tools from Mistral, Phind, Databricks (DBRX),
Perplexity and others claim attention.

As far as mathematical material for testing is concerned, zbMATH and MathSciNet have collected most of the truly mathematical literature since Zbl's start in 1862 and this amounts to several million items.  Importantly, zbMATH Open offers a full API obeying the OAI-PMH standards and does not require any fees.  But arXiv also makes available, through an API, several hundred thousand recent papers, and their abstracts are readily retrievable too.

Obviously, the ultimate intention is to produce an automated device to suggest classifications for new material.  
Developing such a software
system has to start with an idea of what the process can be, and so 
some initial exploration is required to guide further work.
That is the purpose of the current investigation, which reports on initial trials by hand of the basic ideas involved.

We took as a sample the most  recent items
that had been submitted to arXiv, and used  their included
abstracts or summaries as input for prompting dialogues with ChatGPT that elicited that chatbot's suggested classification. The reason for using the most recent arXiv items was that we could then be confident that these items had not yet been used in LLM pre-training sets.  That gave us 56 varied items ranging over all the areas of mathematics. The number 56 resulted from excluding several duplicate items (as explained in Section~\ref{sec:methods}) from the possible 63, resulting from 1 per top-level classification. 
We then evaluated how well the chatbot could
do in comparison with the general classes provided by the author or the editorial process at arXiv; presumably these are human assignments.  We looked at how many of the automatic results compared with
what items were already labeled.  In 22 cases there was a clear discrepancy. For these items, we considered the classifications more carefully, as would have been done at MR, say, and remarked on some ways they may have come about. 
The results of this simple experiment are reported here.

\subsection*{The MSC 2020 Classification System}
The MSC 2020 Mathematics Subject Classification (MSC)~\cite{MSC2020} is a scheme used by the \textit{Math Reviews} and \textit{Zentralblatt Math} article reviewing services.  
The MSC has also been prepared as Linked Open Data \cite{MSC10-LOD} \cite{MSC20-LOD}.

The MSC has about five dozen top-level subject classes, each given by a two-digit code.
Each of these is divided into several second-level area codes, each specified by an added letter.
The special second-level code ``-'' denotes special kinds of materials, for example, reference or expository works, computer programs, proceedings, and so on.
Within each area, a third-level two-digit code may be given to specify a specific subtopic.  
The third-level code ``99'' means ``None of those defined but in this section''.  A strict count gives 63 two-digit classifications, 529 three-digit classification codes, and 6,022 five-digit classification codes.  

MSC 2020 classifications must contain at least the top-level two-digit code.  They may be additionally specified with a second-level letter or in full with a third-level code.   Sometimes a third-level code of ``xx'' or ``XX'' is given for uniformity, \textit{e.g.} writing 30Fxx instead of 30F.

\subsection*{Related Work}
Considerable attention has been given to document classification by various means, including using large language models.  
We do not attempt a complete summary here.    For a general survey spanning traditional techniques to deep learning, see~\cite{LiQian_2022}.   A brief overview is given by~\cite{Ranjan_2023}.
Relating more specifically to mathematics, the article~\cite{SoWatt:Empirical} explores document classification by $N$-gram analysis, and the article~\cite{Scharpf_2023} examines the use of machine learning to discover formula concepts.

More relevant to what we report here is the paper \cite{AutoMSC}
where machine learning is applied to classification in the context of zbMATH.  But this paper develops custom trained ML, and
is not experimenting with a general public tool.  That is also the case for other more recent projects addressing math classification \cite{Scharpf-mining}.

\subsection*{Outline}
The remainder of this article is organized as follows:
Section~\ref{sec:example} provides an initial example using an LLM to classify a mathematical document, demonstrating this approach is worth examining.
Section~\ref{sec:methods} gives the details of how articles in a systematic sample were selected and how a chatbot was used to give them classifications.
Section~\ref{sec:results} shows the results of the experiment, including when the suggested classes matched and when they differed.
Section~\ref{sec:discussion} discusses the quality of the LLM classification when the classes did not match, and possible sources of bias.
Section~\ref{sec:future} lays out possible future directions and Section~\ref{sec:conclusions} summarizes our conclusions.
\begin{figure}[t]
\begin{center}
\begin{minipage}{.625\textwidth}
    \includegraphics[width=\textwidth]{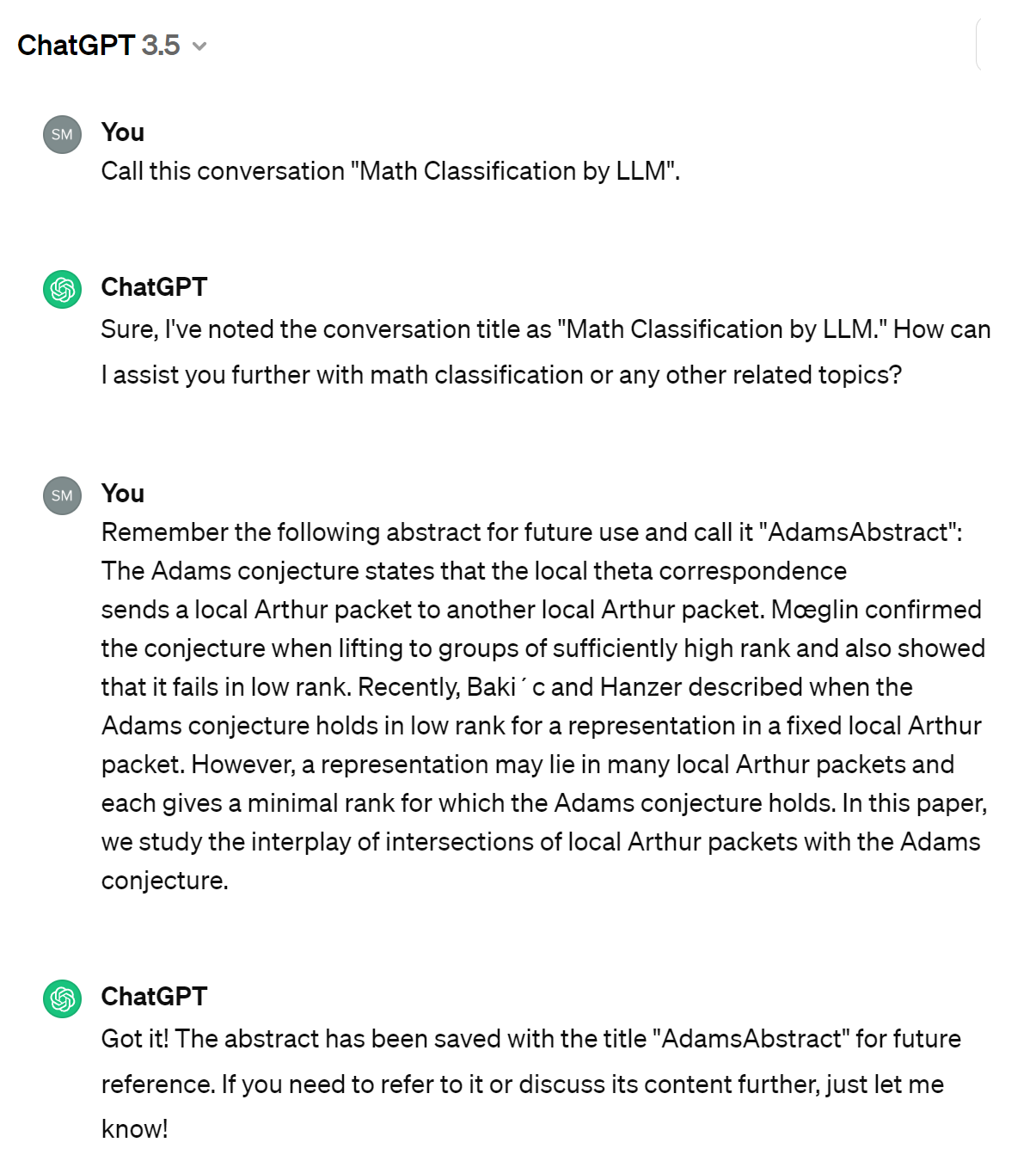}\newline
    \includegraphics[width=\textwidth]{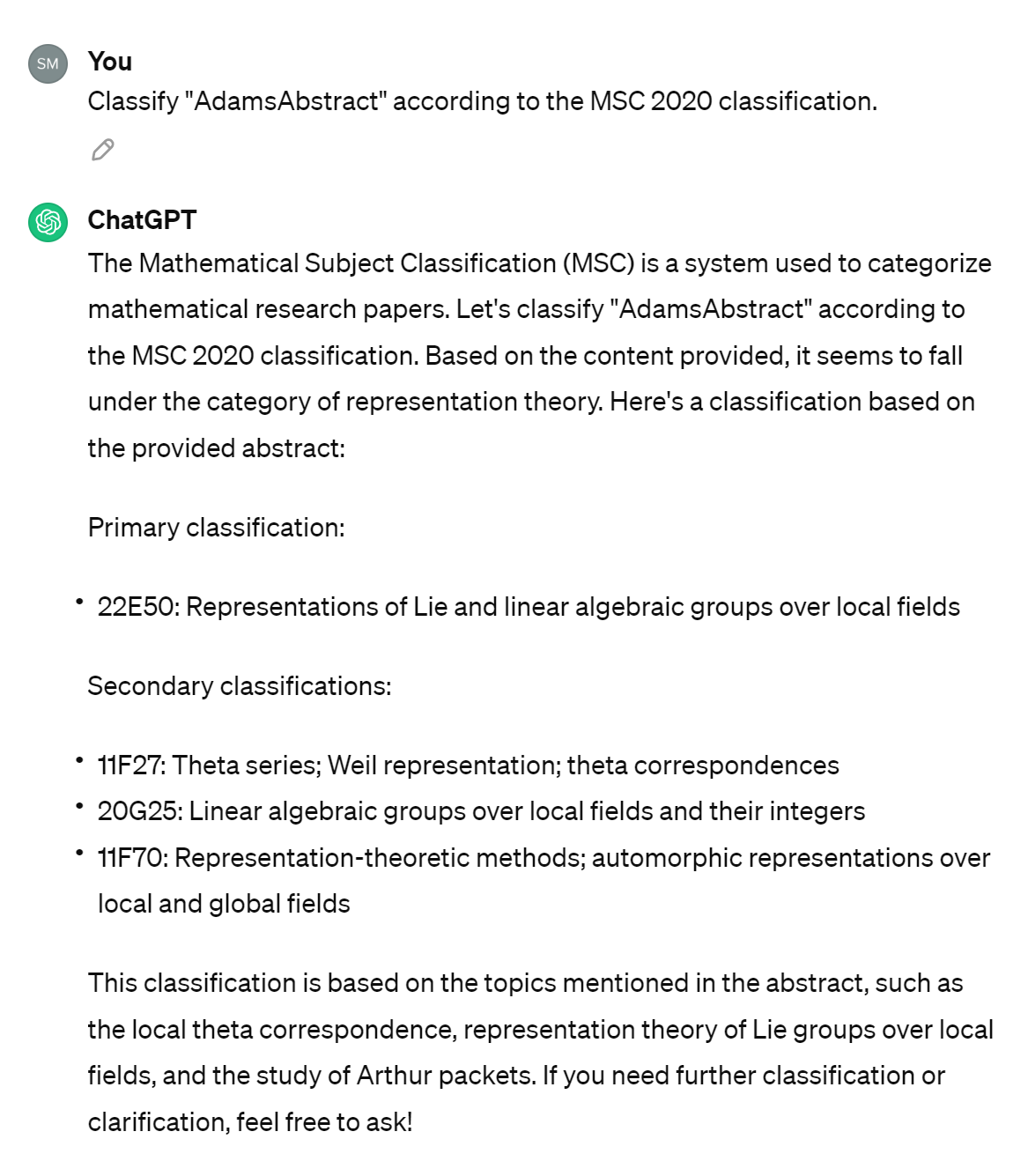}
\end{minipage}
\end{center}
    \caption{Initializing a ChatGPT 3.5 conversation and prompt for an MSC classification}
    \label{fig:Eg1_1}
\end{figure}
\begin{figure}[t]
    \centering
    \includegraphics[width=.75\textwidth]{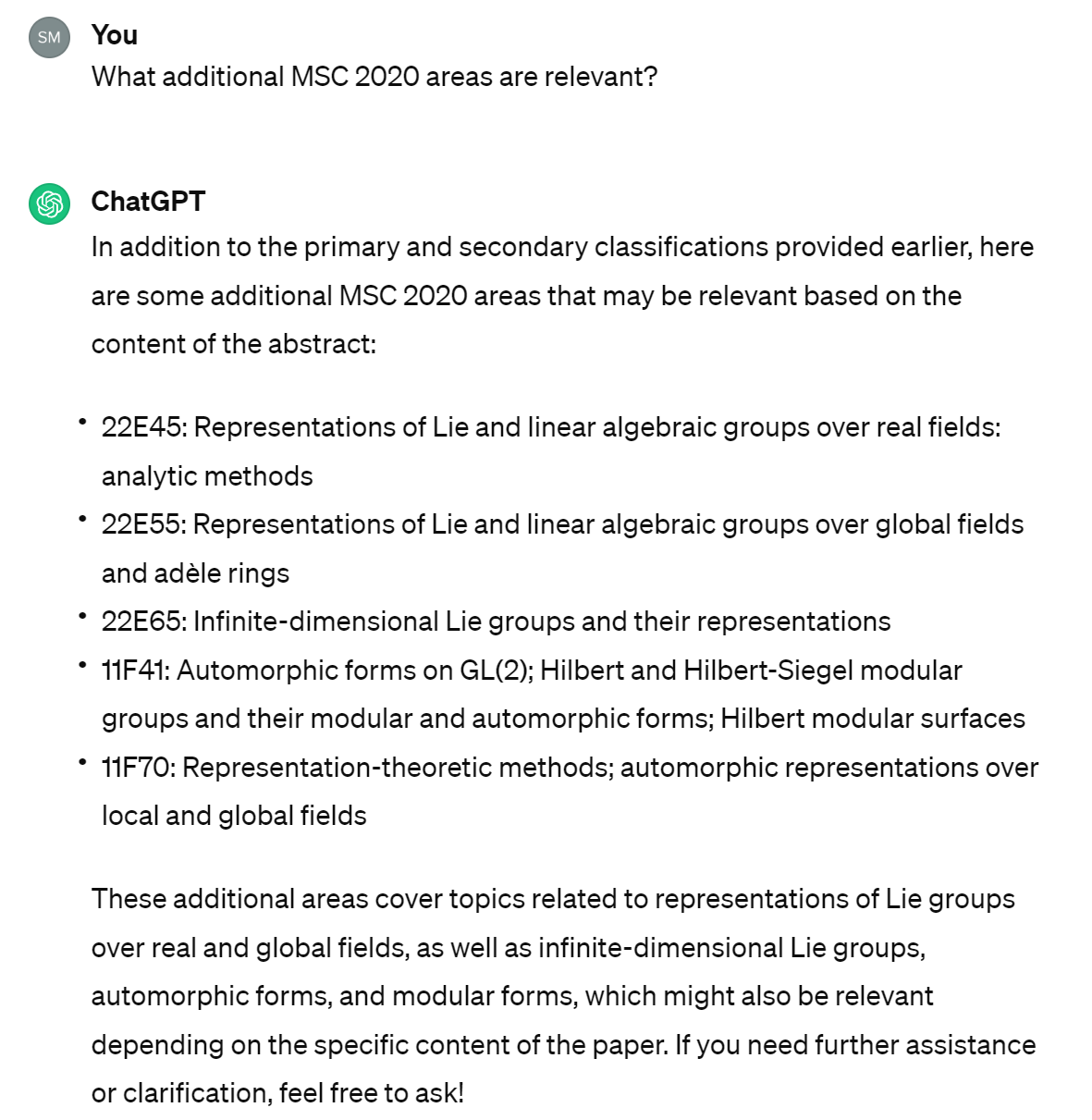}
    \caption{ChatGPT 3.5 prompt to broaden results}
    \label{fig:Eg1_3}
\end{figure}
\section{An Initial Example}
\label{sec:example}
To illustrate the use of LLM tools for document classification, 
we first took as an example an article submitted to the arXiv in the prior 24 hours, namely~\cite{AdamsEg}.  
The author-provided MSC 2020 classes were  11F27, 22E50, 11F70.

The first LLM tool examined was ChatGPT 3.5~\cite{chatgpt3.5} which was freely publicly available.
The text of the abstract was cut and pasted into a conversation, followed by a request to classify it using the MSC 2020 mathematics subject classification system.  This conversation is shown in Figure~\ref{fig:Eg1_1}.   
The primary classification generated by ChatGPT 3.5 was 22E50, and the secondary classifications were 11F27, 20G25, and 11F70.  
All of the author's classes were reproduced, albeit with one additional class; the extra class provided could be considered redundant by an editor at the reviewing organizations but could be argued to provide a suitable hook for a search starting from a different point of view.
An additional prompt asked for additional relevant MSC 2020 areas, as shown in Figure~\ref{fig:Eg1_3}.
\begin{figure}
    \centering
    \includegraphics[width=\textwidth]{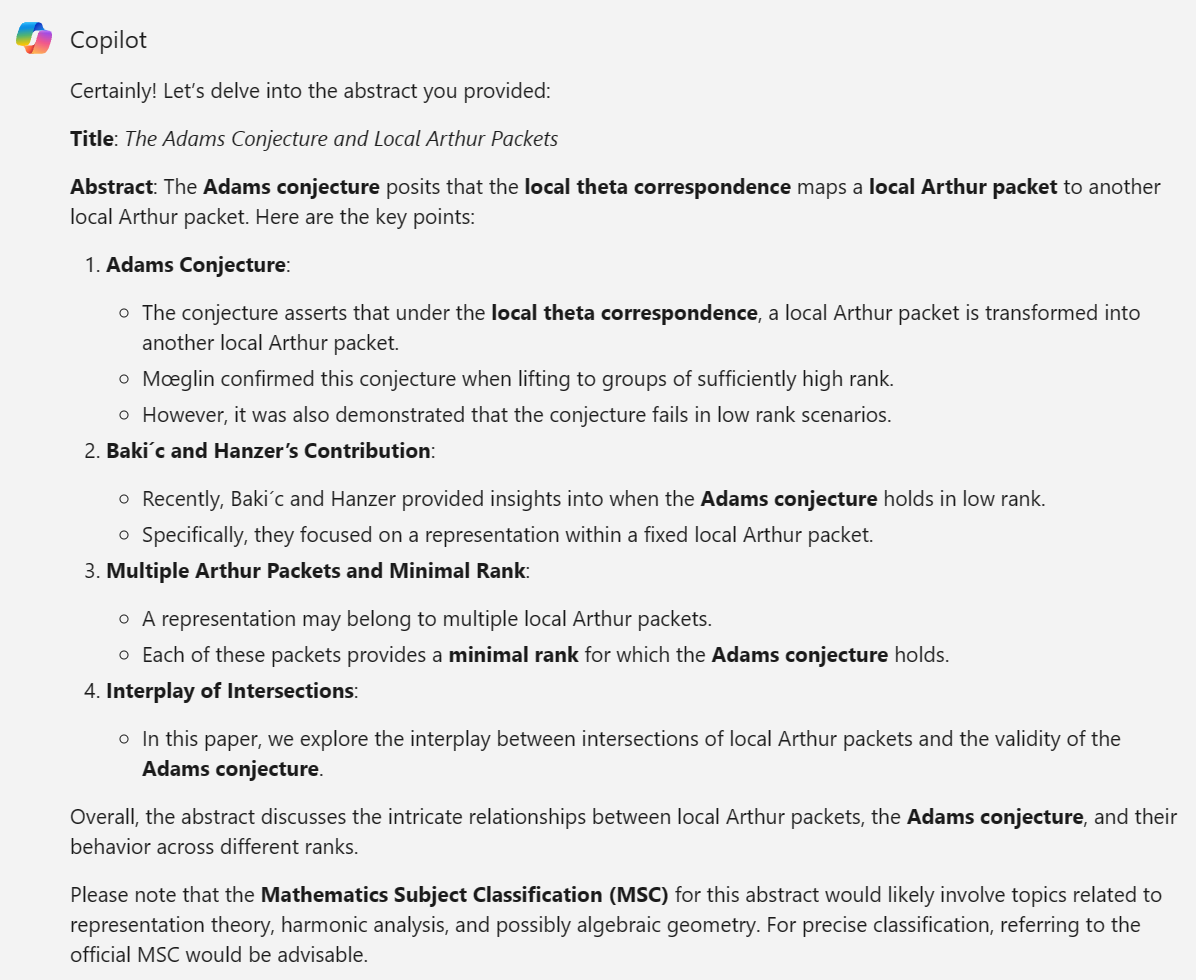}
    \caption{Microsoft Copilot (March 2024) response}
    \label{fig:Eg2_1}
\end{figure}
\begin{figure}
    \centering
    \includegraphics[width=\textwidth]{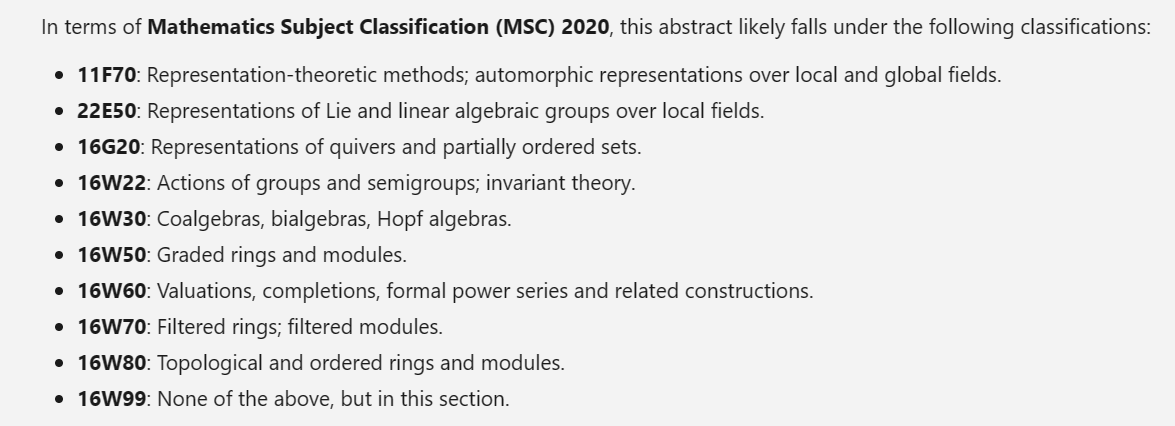}
    \caption{Copilot result for MSC 2020 classification}
    \label{fig:Eg2_2}
\end{figure}

The second tool examined was ChatGPT4, a paid subscription service.
Using the same prompts gave a result saying the most relevant classifications were likely to be 11F70 and 22E50, with 11Fxx and 22E55 as also being potentially relevant depending on the specific content of the paper.

In contrast, Microsoft Copilot (March 2024), when given the following prompt
\begin{verbatim}
    Use the following article abstract for the next prompts:
    <text of abstract> 
\end{verbatim}
responded as shown in Figure~\ref{fig:Eg2_1}.   When further prompted
\begin{verbatim}
    What MSC Subject classifications apply to this?
\end{verbatim}
the same response was given.  
The following, more direct prompt 
\begin{verbatim}
    Give specific subject areas for this abstract using MSC 2020.
\end{verbatim}
repeated the same response but replaced the last paragraph with the specific classifications shown in Figure~\ref{fig:Eg2_2}.  The first two classes, 11F70 and 22E50, were the top classes identified and were two of the three classes provided by the author.  Notably absent was the author's primary classification 11F27.  The other 8 classifications were arguably off-topic.

The alternative Google Gemini (March 2024) produced the classifications 11Fxx and 11S80.

In short, the two versions of ChatGPT and  Microsoft Copilot all identified 22E50 as a suitable classification based on the abstract alone, with 11F70 also suggested at first or in response to a second prompt.  Many of the second prompt responses from Microsoft Copilot and all of the Google Gemini responses were rather broad.  But this experience suggested a systematic sampling could be useful.

\section{Methods}
\label{sec:methods}

A sample of recent research papers was collected from the preprint archive arxiv.org.
One paper was taken from each of the top-level MSC classifications except for the new 97 (Mathematical Education). 
The papers were identified using the arXiv.org MSC search function, searching under ``all fields'' and not just ``mathematics''.
To avoid items already used in LLM training, the most recent paper (as of 2024-04-02) in each of the classifications was taken.
Withdrawn papers were not considered as there was no associated file for analysis. 
Six papers appeared as the most recent item in more than one subject classification. 
The following arXiv items were the most recent in two MSC classifications: 2312.02400, 2312.13673, 2402.07343, 2403.05453, and 2403.16849, while 2311.15913 was the most recent in three.
Each of these was treated only once and contributed a weight of only 1 in the analysis.   
The resulting 56 items used are listed in Tables~\ref{tbl:matching} and~\ref{tbl:differing}.
\begin{table}[!t]
\begin{scriptsize}
\begin{tabular}{l@{~~}l@{~~}l@{~~}ll@{~~}cccc}
\vhead{MSC Section} &
\hhead{arXiv ID} &
\hhead{arXiv MSC} &
\hhead{LLM primary} &
\hhead{LLM secondary} &
\vhead{\# arXiv MSC} & 
\vhead{\# LLM primary MSC} & 
\vhead{\# primary missed} &
\vhead{\# secondary extra} 
\\[1.5ex]
 00 & \arxivid{2403.16849} & (00; 97; 97A99; 97A40) & 00A05 & 97U50                     & 2 & 1 & - & - \\
 01 & \arxivid{2311.16649} & (01)                   & 01Axx & \mcell{01A55, 01A60;}{30Exx,  30Fxx, 30Gxx}
                                                                              & 1 & 1 & - & 1 \\
 03 & \arxivid{2403.07616} & (03)                   & 03Cxx & 03C45, 03C13, 03C98       & 1 & 1 & - & - \\
 05 & \arxivid{2401.13436} & (05; 11)               & 11-xx & 11Pxx, 11Dxx, 11Nxx       & 2 & 1 & 1 & - \\
 08 & \arxivid{2301.09966} & (68; 40; 08)           & 68Q42 & 68Q45                     & 3 & 1 & 2 & - \\
 11 & \arxivid{2403.05453} & (11; 14)               & 14G05, 11G25 &                    & 2 & 2 & - & - \\
 12 & \arxivth{9807008}    & \mcell{(57R70; 58A5; 12; 58C35;}{40 58F19; 58Q15)}
                                          & 58J52 & \mcell{53C05,}{ 58J40, 58J50, 58J52}& 3 & 1 & 2 & 1 \\
 15 & \arxivid{2403.19691} & (15)                   & 15a42, 15A60, 15A90 &             & 1 & 1 & - & - \\
 16 & \arxivid{2401.02545} & (16) 
              & 16-xx & 17-xx 
              & 2 & 1 & - & 1 \\
 18 & \arxivid{2307.01498} & (18)                   & 18D05 & 18-02, 18D35, 18D99       & 1 & 1 & - & - \\
 19 & \arxivid{2112.12010} & (14; 13; 11; 19)       & none  &                           & 4 & 0 & 4 & - \\
 30 & \arxivid{2312.13673} & (30; 31)               & 30C15 & 26C10                     & 2 & 1 & 1 & 1 \\
 32 & \arxivid{2112.13212} & (32; 53)               & 32U05, 32U15 & 32W20, 46G20       & 2 & 1 & 1 & 1 \\
 35 & \arxivid{2311.15913} & (34; 35; 49; 70; 74)   & 49M25 & 49K15                     & 5 & 1 & 4 & - \\
 37 & \arxivid{2403.13116} & (37)                   & 37-xx & 37Hxx                     & 1 & 1 & - & - \\
 43 & \arxivid{1509.03761} & (42; 43; 46)           & 42B25 & 46E35                     & 3 & 1 & 1 & - \\
 45 & \arxivid{2102.03123} & (45; 74; 76; 86)       & 74-xx & 74Fxx                     & 4 & 1 & 3 & - \\
 47 & \arxivid{2312.06390} & (47)                   & 47B37 & 47B15                     & 1 & 1 & - & - \\
 52 & \arxivid{2204.00893} & (90; 05; 52; 68)       & 68T10 & 68W10                     & 4 & 1 & 3 & - \\
 53 & \arxivid{2312.08173} & (70F; 53)              & 70G65 & 83C10                     & 2 & 1 & 1 & 1 \\
 55 & \arxivid{2308.15974} & \mcell{(51; 53; 55; 57K10;}{~58D05; 37E30)}
                                          & 55R65 & 57S30                     & 6 & 1 & 4 & - \\
 58 & \arxivid{2307.00694} & (35; 53; 58)           & 58J50 & 58J32                     & 3 & 1 & 2 & - \\
 62 & \arxivid{2403.18245} & (62)                   & 62P20 & 62H12                     & 1 & 1 & - & - \\
 65 & \arxivid{2403.07875} & (65)                   & 65M12 & 65M55                     & 1 & 1 & - & - \\
 68 & \arxivid{2404.00549} & (68)                   & 68T45 & 92C55                     & 1 & 1 & - & - \\
 76 & \arxivid{2403.18088} & (65 Prim; 76; 35 Sec)  & 76M20 & 65M60                     & 1 & 1 & - & - \\
 78 & \arxivid{2403.08471} & (78)                   & 78A45 & 78A35                     & 1 & 1 & - & - \\
 80 & \arxivid{2309.02308} & (78; 80)               & 78A60 & 78M20                     & 2 & 1 & 1 & - \\
 82 & \arxivid{2312.14281} & (82)                   & 82D45 & 82C26                     & 1 & 1 & - & - \\
 83 & \arxivid{2402.13860} & (83)                   & 83-xx & 83Cxx                     & 1 & 1 & - & - \\
 85 & \arxivid{2401.02337} & (85)                   & 85-xx & 85A04                     & 1 & 1 & - & - \\
 92 & \arxivid{2312.12888} & (90; 92)               & 92B20 & 68T05                     & 2 & 1 & 1 & - \\
 93 & \arxivid{2402.13772} & (93)                   & 93B35 & 93B07                     & 1 & 1 & - & - \\
 94 & \arxivid{2308.14725} & (20; 68; 94)           & 94A60 & 20D60, 20F50, 20P05       & 3 & 1 & 1 & - \\
\end{tabular}
\vspace{\baselineskip}
\end{scriptsize}
\caption{\textbf{Matching Classifications.}
The table shows the 34 cases where all LLM-suggested primary classifications were among the arXiv classifications.
The four right-hand columns are explained in Section~\ref{sec:results}.
}
\label{tbl:matching}
\end{table}
\begin{table}[!ht]
\begin{scriptsize}
\begin{tabular}{l@{~~}l@{~~}l@{~~}l@{~~}l@{~~}ccc@{~~}c@{~~}cc}
\vhead{MSC Section} &
\hhead{arXiv ID} &
\hhead{arXiv MSC} &
\hhead{LLM Primary} &
\hhead{LLM Secondary} &
\vhead{\# arXiv MSC} & 
\vhead{\# LLM primary MSC} & 
\vhead{\# LLM primary ``wrong''} & 
\vhead{\phantom{\#} LLM quality} & 
\vhead{\# primary missed} & 
\vhead{\# secondary extra} 
\\[1.5ex]
 06 & \arxivid{2403.05604} & (06)         & 05C15 & \mcell{05C20, 05C75,}{05C85, 06-xx}& 1 & 1 & 1 & $=$  & - & 1  \\
 13 & \arxivid{2306.17679} & (13; 03)     & 16W10 & 16K20, 16H05              & 2 & 1 & 1 & $+1$ & 2 & 1  \\
 17 & \arxivid{1910.03789} & (17)         & 85-xx & 85Axx                     & 1 & 1 & 1 & $+2$ & 1 & 1  \\
 20 & \arxivid{2308.15765} & (20)         & 94A60 & 11T71                     & 1 & 1 & 1 & $+2$ & 1 & 1  \\
 22 & \arxivid{2303.01437} & (22; 76)     & 35Q72 & 74B20                     & 2 & 1 & 1 & $+2$ & 2 & 1  \\
 26 & \arxivid{2312.02400} & (26; 40)     & 68-xx & 68Txx                     & 2 & 1 & 1 & $+2$ & 2 & -  \\
 28 & \arxivid{1801.04970} & (28)         & 26A39 & 28A12                     & 1 & 1 & 1 & $-2$ & - & -  \\
 33 & \arxivid{2301.05790} & (33; 33Cxx)  & 35S05 & 30E20                     & 1 & 1 & 1 & $=$ & 1 & 1  \\
 34 & \arxivid{2403.06996} & (34; 37; 60; 62) & 68T50 & 68T05                 & 4 & 1 & 1 & $+1$ & 4 & -  \\
 39 & \arxivid{2312.03569} & (39; 81)     & 34A05 & 34L10                     & 2 & 1 & 1 & $+2$ & 2 & 1  \\
 41 & \arxivid{2402.09991} & (41)         & 86-xx & 94A17                     & 1 & 1 & 1 & $+1$ & 1 & 1  \\
 42 & \arxivid{2311.07436} & (42)         & 47B47 & 42B20                     & 1 & 1 & 1 & $=$ & - & -  \\
 44 & \arxivid{2212.04345} & (33; 44; 81) & 42A38 & 81S99                     & 3 & 1 & 1 & $+2$ & 2 & -  \\
 46 & \arxivid{2302.05234} & (46)         & 82B44 & 47B80                     & 1 & 1 & 1 & $+2$ & 1 & 1  \\
 51 & \arxivid{2402.07343} & (14; 51; 81) & 53D12, 58J42 &                    & 3 & 2 & 2 & $+2$ & 3 & 1  \\
 54 & \arxivid{2303.13253} & (70; 97; 54) & 01 & 70                           & 3 & 1 & 1 & $+1$ & 2 & -  \\
 57 & \arxivid{2403.19481} & (57)         & 58J10 & 53C23                     & 1 & 1 & 1 & $+1$ & 1 & 1  \\
 60 & \arxivid{2403.15220} & (60)         & 62-xx, 62Fxx, 62F10 & 68-xx, 68Txx, 68T01
                                                                    & 1 & 1 & 1 & $+2$ & 1 & 1  \\
 74 & \arxivid{2311.17485} & (74)         & 65F90 & 65N99                     & 1 & 1 & 1 & $+2$ & 1 & 1  \\
 86 & \arxivid{2401.06225} & (86)         & 68U10 & 62H30, 68T45              & 1 & 1 & 1 & $+2$ & 1 & -  \\
 90 & \arxivid{2402.12283} & (90)         & 65K05 & 65K10                     & 1 & 1 & 1 & $+2$ & 1 & -  \\
 91 & \arxivid{2402.15849} & (91)
    & 68M10 & 91A80
  & 1 & 1 & 1 & $=$ & - & - \\
\end{tabular}
\vspace{\baselineskip}
\end{scriptsize}
\caption{\textbf{Differing Classifications.}
This table shows the 22 cases where at least one of the LLM-suggested primary classifications was not among the arXiv classifications.
There are two columns that do not appear in Table~\ref{tbl:matching}, namely \textbf{\# primary ``wrong''} and \textbf{LLM quality}, as explained in Section~\ref{sec:discussion}.
}
\label{tbl:differing}
\end{table}

Tables~\ref{tbl:matching} and~\ref{tbl:differing} record the arXiv MSC classification as given with the article summary.
In most cases, the classification was given as a set of 2-digit top-level codes.  
Between one and six codes were provided.   
In 3 cases, a more refined classification (e.g. 33Cxx or 57K10) was given.  In all cases, we used only the top-level 2-digit code in the analysis.
The codes are given in the columns labeled ``arXiv MSC'' in the tables.
In 5 of the 56 papers, the paper itself gave MSC classification(s).  
In all cases, the top-level primary classification in the paper was among the top-level classifications listed by the arXiv.

ChatGPT 3.5 was used to classify the items. A record of the session is available at
\cite{llm3.5-math-classification-1}.  
For each of the items, the title and abstract were provided to the chatbot and an MSC 2020 classification was requested as follows:
\begin{small}
\begin{verbatim}
    Call the following text "<arXivId>-Title": <title from paper>

    Call the followign text "<arxivXivId>-Abstract": <abstract from paper>     

    Given the title "<arXivId>-Title" and abstract "<arXivId>-Abstract" 
    classify the text according to the MSC 2020 classification. 
\end{verbatim}
\end{small}
The title and abstract were moved by cut-and-paste from the PDF file for the paper.  
The quality of the mathematical formula representation varied greatly, from reasonable use of Unicode to a smattering of ASCII approximations to the symbols.  In all cases, the two-dimensional structure of the formula was lost.  
Only minimal editing was done to the text cut-and-pasted, specifically, control codes corresponding to ligatures (\textit{e.g.} ``ffi'') were replaced with their expanded equivalents.
In some cases there was no ``Abstract'' section, \textit{per se}, but there was a part with a similar function, \textit{e.g.} ``Summary'', in which case that was used.  In a small number of cases there was no abstract in the paper, in which case the introduction was used. In all cases, the text used came from the paper itself, not from the arXiv announcement.

In response to the prompt for classification, the chatbot usually gave the primary and secondary classifications separately.
If it did not give secondary classifications, these were requested.   
The resulting primary and secondary classifications are shown in  Tables~\ref{tbl:matching} and~\ref{tbl:differing}.

The chatbot sometimes indicated that the classification ``likely falls'' in certain areas, or ``can be classified'' as what it gave, or with similar cautionary language.  In other cases, it used definitive language.  ChatGPT 3.5 and ChatGPT 4 displayed differences in the specific wording, but since the text generated is the result of a randomizing process it is unclear if anything should be made of this without a great deal more samples.  In one case, arXiv Id 2112.12010, the chatbot gave no specific classification, stating the abstract did not have enough detail:
\begin{quote}
\begin{minipage}{0.9\textwidth}
\begin{footnotesize}
\begin{verbatim}
Without specific details from the abstract, it's challenging
to classify  the text accurately according to the MSC 2020
classification. The  abstract mentions recent developments in
algebraic geometry over p-adically complete commutative rings,
but it lacks specific topics or areas covered in the paper.
For a more precise classification, additional information or
content from the abstract is needed. 
\end{verbatim}
\end{footnotesize}
\end{minipage}
\end{quote}

\section{Results}
\label{sec:results}
In 34 of the 56 cases, 
all LLM-suggested primary classifications were among the arXiv classifications.
These are shown in Table~\ref{tbl:matching}.
In 22 of the 56 cases,
at least one of the LLM-suggested primary classifications was not among the arXiv classifications.
These are shown in Table~\ref{tbl:differing}.
In all cases, it is only the 2-digit top-level code that is considered, even though more specific codes were suggested.  This is because this is all the arXiv-provided MSC classifications had in most cases.

Analyzing the 34 cases where all the LLM-suggested primary classifications were among the arXiv classifications, we see:
In 17 cases, one or more of the arXiv primary classifications were missing from the LLM-suggested primary or secondary classifications.   
In 6 cases, the LLM suggested additional secondary classifications.  
For MSC search key 12, the arXiv classification ``40'' appears to have been a typographical error and was ignored.

Analyzing the 22 cases where at least one of the LLM-suggested primary classifications was not among the arXiv classifications, we see:
In 18 cases one or more of the arXiv classifications were missing from the LLM-suggested primary or secondary classifications.  These are examined in Section~\ref{sec:discussion}.
In 13 cases one of the LLM-suggested additional secondary classifications was neither among the arXiv classifications nor the wrongly LLM-suggested primary classification.
In the next section we discuss in some detail each of the cases where the LLM classification differed from the classification under which the paper was found.   We find that in all but one case, the LLM classification was as good or better.
\section{Discussion}
\label{sec:discussion}
\subsection*{Analysis of Differences}
About 40\% of the cases examined 
showed differing classifications. This can be considered pretty good for a system that was in no way prepared for the task it was performing.  The results are even more surprising when you analyze the discrepancies, as we now do.

We can take a look at the entries where the chatbot gives a ``wrong'' primary two-digit classification code and see ``how wrong'' they are. We give a ``quality score'' on a scale from a scale of -2 to +2, as follows:
\begin{quote}
\begin{enumerate}
\item[$+2$] LLM better than arXiv class
\item[$+1$] LLM slightly better than arXiv class
\item[$=$] arguable either way
\item[$-1$] LLM slightly off
\item[$-2$] LLM way off
\end{enumerate}
\end{quote}
There can be cases when the abstract is inadequate, and anybody would have a hard time doing better.  We list the actual arXiv id, so we can be second-guessed, and the two sets of MSC associated for the 22 items.

\paragraph{arXiv:2403.05604: under 06 LLM: 05C15, 05C20, 05C75, 05C85, 06-xx}

This short paper is certainly about 06A07 ---  Combinatorics of partially ordered sets;
under 06 --- Order, lattices, ordered algebraic structures [See also 18B35];
but then it's also about small finite combinatorial examples and can be argued to touch upon the 05 areas
the chatbot mentions; in any case, the chatbot has 06 anyway.  Score: $=$.

\paragraph{arXiv:2306.17679v1: under 13; 03; LLM: 16W10, 16K20, 16H05}

This is a special case.   The query was based on the paper's Introduction, not on the arXiv abstract that seems to be attached at submission time. The intro mentions ``fields and Brauer algebras'' whereas the abstract had ``etale topology and the notion of Azumaya algebra over a commutative ring ''.  This shows the general difficulty of classification based on small snippets.  13 --- Commutative algebra is not enough.  Azumaya algebras turn up more in 16W and 14. Score: +1.

\paragraph{arXiv:1910.03789 under 17; LLM: 85-XX, 85A-XX    }
 The arXiv's 20 is not appropriate;
The chatbot is quite right. However, the formulation 85A-XX is malformed as an MSC code; it should be 85Axx. Score:+2.

\paragraph{arXiv:2308.15765: under 20 LLM: 11T71, 94A60}

The arXiv's 20 is really not appropriate;
The chatbot is quite right: 11T71 --- Algebraic coding theory; cryptography (number-theoretic aspects)
94A60 --- Cryptography [See also 11T71, 14G50, 68P25, 81P94] Score: +2

\paragraph{arXiv:2303.01437 under 22, 76; LLM: 35Q72	74B20}
MSC 35Q is correct and better; 76 for fluids
does not seem better than 74B for elastic materials (in this case soft solids).
Some sort of 22 for the Lie-group approach seems good as a secondary.  But a caveat in all this is that 35Q72 is a chatbot hallucination: there is no such code in MSC2020! Score: +2 if you ignore the final detail.

\paragraph{arXiv:2312.02400 under 26, 40; LLM: 68-XX, 68Txx}
The chatbot is right though 68 and 68T should be reversed.
26 and 40 don't seem appropriate even considering the references. Score: +2.

\paragraph{arXiv:1801.04970 under 28; LLM: 26A39, 28A12}
MSC 28B05 seems good, as opposed to the single-variable 26 classes,
The author's previous work, including a voluminous book, has been
classified as 60G, 60H, 28A, and 28C.  Here the chatbot is acting superficial.
Score: -2.

\paragraph{arXiv:2301.05790 under 33, 33Cxx; LLM: 35S05, 30E20}
MSC 33 is a little superficial; the chatbot's not good either.
The paper claims to be clarifying material that includes MSC codes
58J05 --- Elliptic equations on manifolds, general theory;
31C12 --- Potential theory on Riemannian manifolds and other spaces;
33C05 --- Classical hypergeometric functions, $_2F_1$;
33C45 --- Orthogonal polynomials and functions of hypergeometric type (Jacobi, Laguerre, Hermite, Askey scheme, etc.);
35A08 --- Fundamental solutions to PDEs;
35J05 --- Laplace operator, Helmholtz equation (reduced wave equation), Poisson equation;
42A16 --- Fourier coefficients, Fourier series of functions with special properties, special Fourier series.  This is better all around, especially 58J; one could argue too for 35L, 35R (manifolds).
Score: $=$.

\paragraph{arXiv:2403.06996 under 34, 37, 60, 62;	LLM: 68T50, 68T05}

arXiv is confusing, except that it correctly has cs{.}AI category assigned; 
68T is certainly right as far as the math is concerned.
This paper, which turns out to be on human creativity, may be worth reading for general reasons.  It is certainly difficult to classify by MSC. Score: +1.

\paragraph{arXiv:2312.03569 under 39, 81; LLM: 34A05, 	34L10}
The chatbot is better; 39 isn't right, though a code 
81 secondary would seem OK. Score: +2.

\paragraph{arXiv:2402.09991 under 41;	LLM: 86-XX, 94A17 }
MSC 41 seems just wrong.
86A10 --- Meteorology and atmospheric physics seems right.
94A17 --- Measures of information, entropy 
seems off. Score: +1.

\paragraph{arXiv:2311.07436 under 42; LLM: 47B47, 42B20}
If the chatbot had the right primary it would be better. But it's not operator theory. 
42B20 --- Singular and oscillatory integrals;
47B47 --- Commutators, derivations, elementary operators, etc.
 Score: $=$.

\paragraph{arXiv:2212.04345 under 44, 33,81;	LLM: 42A38,	81S99}
The chatbot does better with
42A38 --- Fourier and Fourier-Stieltjes transforms and other transforms of Fourier type; 
81SXX --- General quantum mechanics and problems of quantization.  Score: +2.

\paragraph{arXiv:2302.05234 under 46; LLM: 82B44,47B80}
It's not 46 --- Operator Theory, but better the chatbot's 47B80 --- Random linear operators [See also 47H40, 60H25].
82B44 -- Disordered systems (random Ising models, random Schrödinger operators, etc.) in equilibrium statistical mechanics. Score: +2

\paragraph{arXiv:2402.07343 under 51, 14, 81; LLM: 53D12, 58J42}

This first of the authors' series of papers devoted to their project ``Holomorphic Floer Theory" on ``exponential integrals in fínite and infínite dimensions" is wide-ranging, long at 169 pages, and authored by a Fields medalist and another figure of international renown.
ArXiv's Primary 51-XX ---
Geometry {For algebraic geometry, see 14-XX; for differential geometry, see 53-XX} is much too generic.
The chatbot has 53D12 --- Lagrangian submanifolds; Maslov index;
58J42 --- Noncommutative global analysis, and noncommutative residues, are more appropriate. Score: +2.


\paragraph{arXiv:2303.13253 under 54, 70, 97; LLM: 01,70}

The paper is entitled ``What is a degree of freedom?" and addresses that question.  So they agree on 70-XX ---Mechanics of particles and systems, but 54 --- General topology is misplaced, and 97 doesn't contribute much.  Area 01 for History and Biography can be argued as reasonable.  Score: +1.

\paragraph{arXiv:2403.19481 under 57; LLM: 58J10, 53C23}

The chatbot's primary 58J10 --- Differential complexes; elliptic complexes, is better;
the secondary 53C23 --- 53C23 Global geometric and topological methods (à la Gromov); differential geometric analysis on metric spaces --- is a little generic. 
But arXiv's 57 --- Manifolds and cell complexes --- is just too general. Score: +1.

\paragraph{arXiv:2403.15220 under 60; LLM: 52-XX, 62Fxx, 62F10, 68-XX, 68Txx, 68T05}

The chatbot is better; we have econometric models and 62Fxx --- Parametric inference 
62F10 --- Point estimation, 
68T05 --- Learning and adaptive systems.  Score: +2.

\paragraph{arXiv:2311.17485 under 74; LLM: 65F90, 65N99 }

The paper is about  Model Order Reduction techniques, i.e., 65Fxx --- Finite difference methods for boundary value problems involving PDEs  and 65Nxx --- Numerical linear algebra, though the examples are indeed from 74-XX --- Mechanics of deformable solids. However, the chatbot's 65F90 is a hallucination. Score: +2

\paragraph{arXiv:2401.06225 under 86; LLM: 68U10, 62H30, 68T45}

This paper about ``Robust Cloud Suppression and Anomaly Detection'' can be argued to be about the needs of meteorology and thus to do with 86.  But 
we see 68U10 ---
Computing methodologies for image processing, 
62H30 ---
Classification and discrimination; cluster analysis (statistical aspects) [See also 68T10, 91C20]; mixture models, 
68T45 --- Machine vision and scene understanding.
The practical problem of clouds is geophysical but the chatbot classes are 
more for the mathematical matters addressed.  Score: +2.

\paragraph{arXiv:2402.12283 under 90; LLM: 65K05, 65K10}

Again the math aspect is 65 (65K05
--- Numerical mathematical programming methods [See also 90Cxx], 
65K10 --- Numerical optimization and variational techniques [See also 49Mxx, 93-08]) though it can be seen as 
a control problem belonging under 90; but one general 90 is not enough; better could be
90Cxx --- Mathematical programming {For numerical methods, see also 49Mxx, 65Kxx}. Score: +2.

\paragraph{arXiv:2402.15849 under 91; LLM: 68M10, 91A80}

The subject of blockchain theory and its uses is hard to classify with the MSC, which
is perhaps appropriate for a new subject.  Though it can be argued that primary 91 --- 
 Game theory, economics, finance, and other social and behavioral sciences, covers
the intent of blockchains, the work here certainly does qualify  more generally as 68M10 --- Network design and communication in computer systems [See also 68R10, 90B18], and we have
91A80 --- Applications of game theory 
Score: =.

\subsection*{Possible Sources of Bias}

The activity of classification, or categorization of any sort, is usually subject to many potential sources of bias.  Here we have, first of all, the small initial sample size reported.  That will be mitigated with
a script to try many more cases; the discrepancies will still have to be surveyed by humans.  It should then be possible also to try both abstracts and bigger input text samples, possibly not collected by cut-and-paste from PDF files.

The articles posted to our source, arXiv, are unrefereed and we don't know how they come to be arranged under MSC sections as they are posted.  The paper authors are not necessarily well familiar with the MSC, and some papers simply don't fit well.
Some authors give many many classes, others give none or one.

\section{Future Directions}
\label{sec:future}
Some of what to do next is clear.  We need to bring to maturity 
script programs that carry out the simple manipulations described in our hands-on test.  Modules for the several public APIs on offer need to be made.

A significant question is whether it is possible to use the sorts of support offered to develop one's own smaller language models, for the sake of independence from unknowable cloud software systems. We do have more readily available sources of MSC-classified ground truth.  It's not clear if this sort of development will work for the special subject of mathematics.

An obvious improvement for a systematic study would be to treat mathematical formulae with care, perhaps using the approach of  \cite{SoWatt:Empirical}.

 A well-trained chatbot that suggests MSC classifications could be a web service offered by zbMATH and MathSciNet to pre-classify materials before submission to journals or arXiv.
 This would be similar to using zbMATCH to get 
 an article's
 bibliography correct.

\section{Conclusions}
\label{sec:conclusions}

This is a very first step to evaluate the viability of using LLMs to classify mathematical material.  A tool in no way customized for mathematics nonetheless provided mechanical classification of our specialized subject material.  
At the same time, it is necessary to watch out for occasional surprising hallucinations, as we have seen twice over MSC in our experiment.  However, it has elsewhere been remarked that ChatGPT can be quite convincing and quite wrong on scientific facts, including even simple arithmetic~\cite{Wolfram-CGPT}.

We have found that ChatGPT3.5 in the majority of cases (about 60\% of our sample), from the title and abstract alone,  never gives a wrong primary subject classification.   At the same time, in half of these ``correct'' cases, it did not deduce all the primary classifications.  This may be a reflection of the level of detail in the abstract.

We have also found that in the minority of cases (about 40\% of our sample), the LLM-suggested primary classification(s) did not match the general classification provided in arXiv.  
Upon examination, however, we have seen that the LLM-suggested primary classifications were actually substantially \textit{better} than the ones provided.

We therefore conclude that the use of LLMs for the classification of mathematical articles is likely to be fruitful and merits further study.

\begin{credits}


\subsubsection{\discintname} The authors have no conflicts of interest to declare.
\end{credits}

\newpage
\IfFileExists{IfExistsUseBBL.tex}{%

}{%
\bibliographystyle{splncs04}
\bibliography{references.bib}
}
\end{document}